# Frequency Dependence of Magnetoelectric Interactions in Layered Structures of Ferromagnetic Alloys and Piezoelectric Oxides

U. Laletsin and N. Padubnaya

Institute of technical acoustics, National Academy of Sciences of Belarus, 21707 Vitebsk, Belarus

G. Srinivasan and C. P. DeVreugd

Physics Department, Oakland University, Rochester, Michigan 48309, USA

ABSTRACT

Magnetoelectric (ME) interactions in layered structures of magnetostrictive and piezoelectric phases are mediated by mechanical deformation. Here we discuss the frequency dependence of ME coupling in bilayers and trilayers of Permendur, a ferromagnetic alloy, and lead zirconate titanate. Data on ME voltage coefficient versus frequency profiles reveal a giant ME coupling at electromechanical resonance. The maximum voltage coefficient of 90 V/cm Oe is three orders of magnitude higher than low-frequency values. The ME interactions for transverse fields is an order of magnitude stronger than for longitudinal fields. These results are in agreement with theory. The resonance ME effect, therefore, is a novel tool for enhancing the magnetic-to-electric field conversion efficiency in the composites.

## 1. Introduction

Magnetoelectric materials facilitate the conversion between energies stored in magnetic and electric fields. In a single phase material, the effect requires the presence of both long-range magnetic order and permanent electric dipole moment [1]. A few single-phase materials are magnetoelectric, but the effect is usually weak [2]. A much stronger magnetoelectric (ME) effect could be realized in a composite of ferromagnetic and ferroelectric phases in which the ME coupling is mediated by mechanical stress. An applied ac magnetic field $\delta H$ produces dynamic deformation due to magnetostriction and results in an electric field $\delta E$ due to piezoelectric effect. The induced polarization $\delta P$ is related to $\delta H$ by $\delta P = \alpha \, \delta H$, where $\alpha$ is the ME-susceptibility. The parameter of importance is the ME voltage coefficient $\alpha_E = \delta E/\delta H$ and is related to $\alpha$ by the expression $\alpha = \varepsilon_o \varepsilon_r \alpha_E$, where $\varepsilon_r$ is the relative permittivity of the material. The ME composites of interest in the past were bulk samples of ferrites with $BaTiO_3$ or lead zirconate titanate (PZT) [3-5]. Bulk composites in general show ME coupling much smaller than predicted values due to leakage currents associated with low resistivity for ferrites. The problems could easily be eliminated in a layered structure since high electrical resistivity leads to the absence of leakage current [6]. The layered structures are also easy to pole in an electric field, thereby strengthening the piezoelectric and ME effect. Our recent studies on bilayers and multilayers of ferrite-PZT and lanthanum manganite-PZT show evidence for ME coupling much stronger than in bulk samples [7,8].

Here we discuss a unique and novel technique for further enhancing the field conversion efficiency of the composite, i.e., ME effects at electromechanical resonance (EMR). This work constitutes the first data for the phenomenon in ferromagnetic alloy-lead zirconate titanate (PZT) layered samples. The resonance ME effect is similar in nature to the standard effect, i.e. an induced polarization under the action of an ac magnetic field. But the ac field here is tuned to the electromechanical resonance frequency. As the dynamic magnetostriction is responsible for the electromagnetic coupling, EMR leads to significant increasing in the ME voltage coefficient. The technique was implemented once in the past for bulk composites in order to distinguish weak ME signals from noise [3].

## 2. Experiment

Bilayer and trilayer structures of ferromagnetic Permendur and piezoelectric PZT were synthesized. Permendur (P) is a soft magnetic alloy consisting of 49% iron, 49% cobalt and 2% vanadium. It is an ideal material for studies on ME composites due to desirable low resistivity and high magnetization (23.4 kG), Curie temperature (1213 K), permeability and magnetostriction (70 ppm) [9]. PZT was chosen due to high ferroelectric Curie temperature and piezoelectric coupling constant [10]. Discs of Permendur and PZT, 9 mm in diameter, were bonded with 0.01-0.03 mm thick layer of an epoxy to fabricate bilayers. Trilayers were made with the central PZT layer bonded to outer P-layers. The thickness of P-layer was 0.18 mm for all the samples studied. PZT thickness varied from 0.2 mm to 0.8 mm. The samples were poled by first heating it to 420 K and then cooling it back to room temperature in an electric field of 5 kV/mm applied perpendicular to the sample plane.

For ME characterization, we measured the electric field produced by an alternating magnetic field applied to the biased composite. The samples were placed in a shielded 3-terminal holder and placed between the pole pieces of an electromagnet that was used to apply the bias magnetic field H. The required ac magnetic field of $\delta H = 0.01-1$ Oe at 10 Hz - 1 MHz parallel to H was generated with a pair of Helmholtz coils. The induced electric field $\delta E$ perpendicular to the sample plane was estimated from the voltage $\delta V$ measured with an oscilloscope. The ME voltage coefficient is estimated from $\alpha_E = \delta E/\delta H = \delta V/t \delta H$ where t is the thickness of PZT. The measurements were done for two different field orientations. With the sample plane represented by (*1,2*), the transverse coefficient $\alpha_{E,31}$ was measured for the magnetic fields H and $\delta H$ along direction-*1* (parallel to the sample plane) and perpendicular to $\delta E$ (direction-*3*). The longitudinal coefficient $\alpha_{E,33}$ is measured for all the fields perpendicular to the sample plane. Magnetoelectric characterization was carried out at room temperature as a function of frequency of the ac magnetic field and bias magnetic field H.

### 3. Results and Discussion

Figure 1 shows representative data on the H dependence of the longitudinal and transverse ME voltage coefficients. The measurements were done at 1 kHz on a trilayer sample with a PZT thickness of 0.36 mm. Data are shown for both increasing and decreasing H. For the longitudinal fields, as H is increased from zero, one observes increase in $\alpha_{E,33}$ to a maximum at $H_m = 600$ Oe. With further increase in H, $\alpha_{E,33}$ decreases rapidly to a minimum. Upon decreasing H from 1.5 kOe to zero, the data show evidence for hysteresis and remanence. When H is reversed, $\alpha_{E,33}$ becomes negative (a 180 deg. phase difference between $\delta H$ and $\delta E$), but the field dependence remains the same as for positive H. For transverse fields, Fig.1 shows a similar variation in the ME voltage coefficient with H as for the longitudinal case but with the following departures. (i) The maximum in $\alpha_{E,31}$ occurs at a much smaller $H_m$ of 150 Oe compared to the longitudinal case. (ii) The peak value of $\alpha_{E,31}$ is an order of magnitude higher than for $\alpha_{E,33}$.

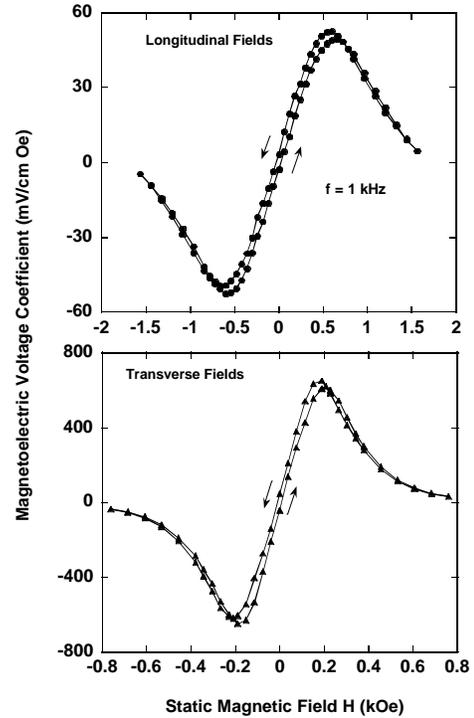

Fig. 1

Fig.1: Variations in longitudinal and transverse magnetoelectric (ME) voltage coefficients with the bias magnetic field H for a trilayer composite of Permendur-PZT-Permendur. The thickness of Permendur (P) and PZT layers are 0.18 mm and 0.36 mm, respectively. The transverse coefficient $\alpha_{E,31} = \delta E_3/\delta H_1$ corresponds to H and an ac field $\delta H$ parallel to each other and to the sample plane (*1,2*) and the induced electric field $\delta E$ measured along the direction-*3*, perpendicular to sample plane. The longitudinal coefficient $\alpha_{E,33} = \delta E_3/\delta H_3$ is for all the fields parallel to each other and along direction-*3*. The data at room temperature and a frequency of 1 kHz are for increasing and decreasing H. The lines are guide to the eye.

Similar measurements on bilayers of Permendur-PZT indicated the presence of strong ME coupling. The coupling strength, however, was a factor of 2.5 smaller than for the trilayers. Next we compare $\alpha_E$ values in Fig.1 with results for similar composites. Systems of interest in the past were bulk samples of ferrites with barium titanate or PZT. The $\alpha_E$ in Fig.1 are a lot higher than values reported for bulk composites of cobalt ferrite (CFO) or nickel ferrite (NFO) with PZT or $BaTiO_3$ [4-6]. Layered composites studied so far include ferrite-PZT [7,8] lanthanum manganite-PZT [11] and terfenol-PZT [12]. For comparison, the highest value for $\alpha_{E,31}$ is 60 mV/cm Oe in bilayers of lanthanum manganites-PZT, 400 mV/cm Oe for NFO–PZT and 4680 mV/cm Oe for Terfenol-PZT. Thus the results in Fig.1 provide clear evidence for one of the highest ME coupling reported for any composites. The field orientation and H dependence of $\alpha_E$ in Fig.1 are in good agreement with theory as discussed later.

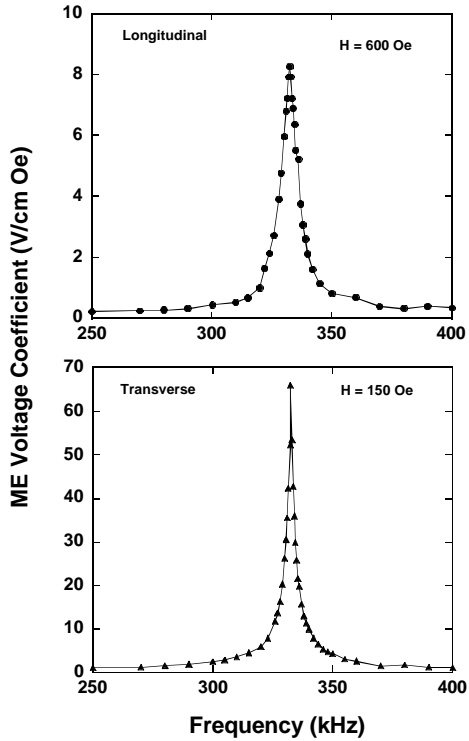

Fig.2: Frequency dependence of transverse and longitudinal ME voltage coefficients for the P-PZT-P trilayer for 0.18 mm thick P-layers and 0.36 mm thick PZT. The bias field H was set for maximum ME coupling (Fig.1). The lines are guide to the eye. The resonance frequency corresponds to the electromechanical (EMR) for PZT.

We also performed studies on the frequency dependence of the ME coupling. The bias field was set at $H_m$ and the voltage coefficients were measured as the frequency f of the ac field $\delta H$ was varied. Typical $\alpha_E$ vs f profiles for longitudinal and transverse fields are shown in Fig.2. The results are for the trilayer with a PZT thickness of 0.36 mm. Consider first the data for the longitudinal field. Upon increasing f, $\alpha_{E,33}$ remains small and constant for frequencies up to 250 kHz. At higher f, we observe a rapid increase in $\alpha_{E,33}$ to a maximum of 8000 mV/cm Oe at 330 kHz. Finally, $\alpha_{E,33}$ levels off at 50 mV/cm Oe at high frequencies. The profile thus shows resonance with $f_r$ = 330 kHz and a width $\Delta f$ = 6 kHz, corresponding to a quality factor Q = 55. Figure 2 shows a similar resonance in $\alpha_{E,31}$ for transverse fields. The resonance occurs at the same frequency as for the longitudinal fields, but with a much higher maximum $\alpha_E$ (=66000 mV/cm) Oe and a higher Q (=155) compared to the longitudinal fields. The transverse coupling is an order of magnitude stronger than the longitudinal voltage coefficient, similar to the observation in Fig.1 for low frequencies.

We carried out measurements of resonance ME effect in trilayers with PZT thickness ranging from 0.2 to 0.8 mm. Figure 3 shows the transverse ME voltage coefficient at resonance as a function of the thickness t of PZT. An initial increase in $\alpha_{E,31}$ with t is evident in Fig.3; it shows a maximum value of 90 V/cm Oe for a PZT thickness of 0.6 mm and then decreases for higher t. The transverse ME voltage coefficient in Fig.3 is the largest ever reported for any composite. In the discussion to follow in this section, we attribute the resonance in $\alpha_{E,31}$ to electromechanical resonance in PZT.

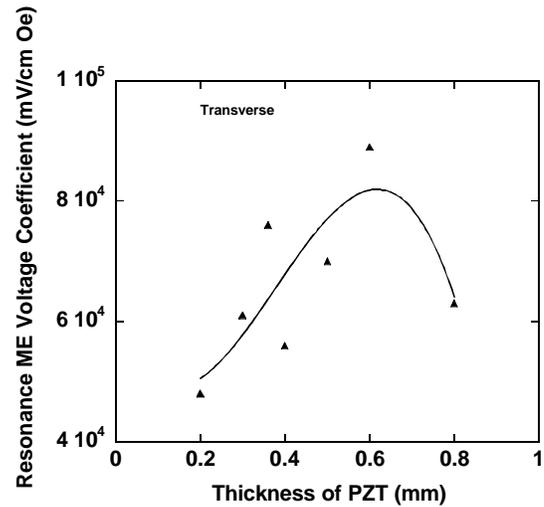

Fig.3: Variation of $\alpha_{E,31}$ at resonance as a function of the thickness of PZT. The lines are guide to the eye.

Now we analyze the data in terms of theoretical models we developed recently for low frequency and resonance ME effects [13,14]. The low frequency data in Fig.1 is considered first. Important observations that need to understood are: (i) the overall H dependence of $\alpha_E$ and (ii) the relative strengths of the ME coupling for longitudinal and transverse field orientations. According to our model for a bilayer, $\alpha_E$ is directly proportional to the product of the piezoelectric coupling constant $d_{31}$ and the piezomagnetic coupling $q=d\lambda/dH$ where $\lambda$ is the magnetostriction [13]. The H dependence in Fig.1 essentially tracks the strength of piezomagnetic coupling. One expects a peak in $\alpha_E$ for maximum in $q$. The piezomagnetic and ME couplings vanish when magnetostriction attains saturation. The transverse coupling is the stronger than the longitudinal coupling due to (i) relative strengths of the piezomagnetic coupling and (ii) demagnetization effects associated with the longitudinal fields. The coefficients $\alpha_{E,31}$ and $\alpha_{E,33}$ are proportional to $d_{13}$ $(q_{11} + q_{12})$ and $d_{13} q_{13}$, respectively. The in-plane parallel piezomagnetic coupling $q_{11}$ is stronger than the out-of-plane term $q_{13}$. A further reduction in $\alpha_{E,33}$ occurs when magnetic fields applied perpendicular to the sample plane are diminished due to demagnetization.

Data on resonance ME effect is considered next. Our recently developed model predicts giant magnetoelectric interactions in ferromagnet-PZT bilayers at frequencies corresponding to electromechanical resonance [14]. We considered a bilayer in the form of thin disk of radius $R$. The ac magnetic field induces harmonic waves in the radial or thickness modes. The model considers radial modes for transverse or longitudinal fields. An averaging procedure was employed to obtain the composite parameters and the ME voltage coefficient $\alpha_E$. The frequency dependence of $\alpha_E$ shows a resonance character at the electromechanical resonance for PZT in the bilayer. The resonance frequency depends on $R$, mechanical compliances, density and the coefficient of electromechanical coupling for radial mode. The peak value of $\alpha_E$ and the width of resonance are determined by the effective piezomagnetic and piezoelectric coefficients, compliances, permittivity and loss factor. Based the model, one expects a resonance in $\alpha_E$ versus frequency profile with a maximum $\alpha_E$ that is a factor of 40-1000 higher than low frequency values, depending on the nature of the magnetostrictive phase.

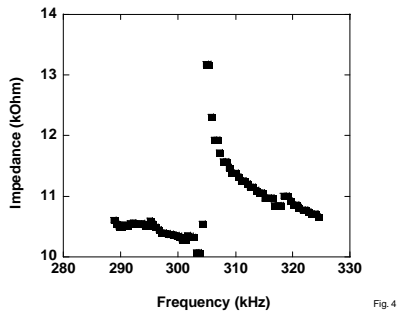

Fig.4: Data on impedance versus frequency showing electromechanical resonance in a poled 9 mm disk of PZT.

Electromechanical resonance (EMR) in composites consisting of piezoelectric phases is usually investigated through measurements of impedence $z$ or dielectric constant as a function of frequency. Figure 4 shows such data for a disc of PZT used in the present study. The results are for a poled 10 mm-diameter sample. One observes EMR with at a frequency of 330 kHz. Thus the resonance frequency for $\alpha_E$ vs f in Fig.2 is in excellent agreement with the expected EMR in PZT. In conclusion, we reported the observation of theoretically predicted giant ME interactions at electromechanical resonance in ferromagnetic alloy-PZT layered samples. The phenomenon could be utilized to accomplish very high field conversion efficiency in the product property composites.

This work at Oakland University is supported by the National Science Foundation (DMR-0302254).

**References**
1. L. D. Landau and E. M. Lifshitz, *Electrodynamics of Continuous Media*, Pergamon Press, Oxford (1960) p.119 (Translation of Russian Edition, 1958)
2. D. N. Astrov, Soviet phys. JETP **13**, 729 (1961).
3. Van Suchtelen, Philips Res. Rep., **27**, 28 (1972).
4. J. Van den Boomgaard, D. R. Terrell, and R. A. J. Born, J. Mater. Sci. **9**, 1705 (1974).
5. J. Van den Boomgaard, A. M. J. G. van Run, and J. van Suchtelen, Ferroelectrics **14**, 727 (1976).
6. G. Harshe, J. P. Dougherty and R. E. Newnham, Int. J. Appl. Electromag. Mater. **4**, 145 (1993); M. Avellaneda and G. Harshe, J. Intell. Mater. Sys. Struc. **5**, 501 (1994).
7. G. Srinivasan, E. T. Rasmussen, J. Gallegos, R. Srinivasan, Yu. I. Bokhan, and V. M. Laletin, Phys. Rev. B **64**, 214408 (2001).
8. G. Srinivasan, E. T. Rasmussen, and R. Hayes, Phys. Rev. B **67**, 014418 (2003).
9. *Ferromagnetism*, R. Bozorth, IEEE Press (New York), 1993.
10. PZT used in the study: sample No.APC850, American Piezo Ceramics, Inc., Mackeyville, Pennsylvania.
11. G. Srinivasan, E. T. Rasmussen, B. J. Levin, and R. Hayes, Phys. Rev. B **65**, 134402 (2002).
12. J. Ryu, A. V. Carazo, K. Uchino, and H. Kim, Jpn. J. Appl. Phys. **40**, 4948 (2001).
13. M. I. Bichurin, V. M. Petrov, and G. Srinivasan, J. Appl. Phys. **92**, 7681 (2002).
14. M. I. Bichurin, D. A. Fillipov, V. M. Petrov, U. Laletsin, and G. Srinivasan, unpublished.